\documentstyle[12pt,a4,epsfig]{article}
\begin{document}
\title{ \vspace*{-3cm}
Identification of baryon resonances in central heavy-ion
collisions at energies between \\ $1$ and $2$ AGeV
}
\author{
M. Eskef$^{6,}$\thanks{on leave of absence from Energy Research Group,
Univerity of Damascus, Syria}, D. Pelte$^6$, G. Goebels$^6$, E. H\"afele$^6$,
N. Herrmann$^{4,6}$, \and
M. Korolija$^{6,12}$,  Y. Leifels$^{4,6}$,
H. Merlitz$^{4,6}$, S. Mohren$^6$, \and
M.R. Stockmeier$^6$, M. Trzaska$^6$, J.P. Alard$^3$, A. Andronic$^1$, \and
R. Averbeck$^4$, Z. Basrak$^{12}$,
N. Bastid$^3$, I. Belyaev$^7$, D. Best$^4$, \and
A. Buta$^1$, R. \v{C}aplar$^{12}$,
N. Cindro$^{12}$, J.P. Coffin$^{10}$, P. Crochet$^{10}$, \and
P. Dupieux$^3$, M. D\v{z}elalija$^{12}$, L. Fraysse$^3$, Z. Fodor$^2$, \and
A. Genoux-Lubain$^3$, A. Gobbi$^4$, K.D. Hildenbrand$^4$, B. Hong$^9$, \and
F. Jundt$^{10}$, J. Kecskemeti$^2$, M. Kirejczyk$^{4,11}$, R. Kotte$^5$,
\and R. Kutsche$^4$, A. Lebedev$^7$, V. Manko$^8$, J. M\"osner$^5$,
D. Moisa$^1$, \and W. Neubert$^5$, C. Plettner$^5$, P. Pras$^3$,
F. Rami$^{10}$, V. Ramillien$^3$, \and W. Reisdorf$^4$, J.L. Ritman$^4$,
B. de Schauenburg$^{10}$, D. Sch\"ull$^4$, \and
Z. Seres$^2$, B. Sikora$^{11}$, V. Simion$^1$, K. Siwek-Wilczynska$^{11}$, \and
V. Smolyankin$^7$, M.A. Vasiliev$^8$, P. Wagner$^{10}$, G.S. Wang$^4$, \and
K. Wisniewski$^{4,11}$, D. Wohlfarth$^5$,
A. Zhilin$^7$ \\[5mm]
The $FOPI$ Collaboration}
\maketitle
\small \noindent \vspace*{-0.5cm}
\newcounter{fig}
\begin{list}{$^{\arabic{fig}}$}{\usecounter{fig}
\labelwidth0.5cm \leftmargin1cm \labelsep0.4cm \rightmargin0cm
\parsep0.5ex plus0.2ex minus0.1ex \itemsep0ex plus0.2ex \topsep0cm
\parskip0cm \partopsep0cm \small }
\item National Institute for Physics and Nuclear Engineering,
Bucharest, Romania
\item Central Research Institute for Physics, Budapest, Hungary
\item Laboratoire de Physique Corpusculaire, IN2P3/CNRS, and Universit\'e
Blaise Pascal, Clermont-Ferrand, France
\item Gesellschaft f\"ur Schwerionenforschung, Darmstadt, Germany
\item Forschungszentrum Rossendorf, Dresden, Germany
\item Physikalisches Institut der Universit\"at Heidelberg, Heidelberg,
Germany
\item Institute for Theoretical and Experimental Physics, Moscow, Russia
\item Kurchatov Institute,  Moscow, Russia
\item Korea University, Seoul, Korea
\item Institut de Recherches Subatomiques, IN2P3/CNRS, and Universit\'e
Louis Pasteur, Strasbourg, France
\item Institute of Experimental Physics, Warsaw University , Poland
\item Rudjer Boskovic Institute, Zagreb, Croatia
\end{list}
\newpage \normalsize
\begin{abstract}
The mass distributions of baryon resonances populated in near-central
collisions of Au on Au and Ni on Ni are deduced by defolding the $p_{\rm t}$
spectra of charged pions by a method which does not depend on a specific 
resonance shape. In addition the mass distributions of resonances 
are obtained from 
the invariant masses of $(p, \pi^{\pm})$ pairs. With both methods the 
deduced mass distributions are shifted by an average value of $-60$ MeV/c$^2$
relative to the mass distribution of the free $\Delta(1232)$ resonance, the
distributions descent almost exponentially towards mass values of $2000$
MeV/c$^2$. The observed differences between $(p, \pi^-)$ and $(p, \pi^+)$
pairs indicate a contribution of isospin $I = 1/2$ resonances. The attempt 
to consistently describe the deduced mass distributions and the 
reconstructed kinetic energy spectra of the resonances leads to new insights 
about the freeze out conditions, i.e. to rather low temperatures and large 
expansion velocities.
\end{abstract}
\section{Introduction}
In central heavy ion collisions the number of mesons produced increases with 
bombarding energy. In the energy range between $1$ and $2$ AGeV, 8\% to 
22\% of the available energy is dissipated into this channel which 
predominantly consists of pions \cite{fopi1,fopi2}. The amount of 
dissipated energy also depends on system mass:
In central Au + Au reactions at $1$ AGeV it is 8\%, but 15\% in
central Ni + Ni reactions at the same bombarding energy. The process
responsible for meson production is believed to be predominantly the
excitation of baryon resonances during the early compression phase of the
collision. In the later expansion phase these resonances decay. The
primordial decay yields a number of different mesons, mainly pions.
Mesons with
masses higher than the pion mass, e.g. $\eta, \rho$, may subsequently
decay also into several pions. In total, the average mass of the excited
baryon resonances and the number of pions produced by their decay chains
increase with bombarding energy. This mechanism is the basic process of pion
production which is used in nuclear transport models to describe the dynamics
of relativistic heavy ion collisions 
\cite{bert88,cass88,cass90,aich91,bass93,bass94,bass95,dani95,teis97a}. 
In a recent publication \cite{teis97a} it was shown that all
resonances up to the $\Delta(1950)$ resonance have to be included to
successfully reproduce the experimental pion data at energies between
1 and 2 AGeV.

The experimental confirmation that baryon resonan\-ces are excited in
relativistic heavy ion collisions is rather scarce. The earliest indication
came from the comparison by Brockmann et al. \cite{broc84} of the pion and
proton energy spectra observed in $1.8$ AGeV Ar $+$ KCl reactions. This
comparison was later extended with the attempt to determine the
$\Delta(1232)$ resonance mass and temperature \cite{sand85}. Similar analyses
of the pion energy spectra measured in $1$ AGeV Au + Au collisions were
published in \cite{seng94}. At much higher energies, i.e. $13.7$ AGeV, the
fraction of nucleons excited to the $\Delta(1232)$ resonance was determined
for Si $+$ Al, Pb reactions \cite{e814}. In this study also the invariant
mass distribution of the $\Delta(1232)$ resonance was reconstructed from
correlated $(p, \pi^+)$ pairs.

To identify structures in the invariant mass distribution of correlated
proton and pion pairs provides the direct
proof that nucleons are excited to high-lying resonances. The major obstacle
in reconstructing the invariant mass is the large background of
non-correlated $(p, \pi)$ pairs, which increases with the number of
protons and is therefore particularly large for central reactions of
heavy-mass systems. In peripheral reactions with very light projectiles,
e.g. p \cite{trza91,chib91} or $^3$He \cite{henn92} induced reactions at
around $2$ GeV bombarding energy, the $(p, \pi^+)$
correlations were successfully analyzed and the mass distribution of the
$\Delta(1232)$ resonance was determined. The resonance mass was found to be
shifted by $-25$ MeV/c$^2$ to lower masses in reactions on various targets,
compared to those on protons. Very recently the $\Delta(1232)$ was
reconstructed from $(p, \pi^+)$ pairs detected at $1.97$ AGeV bombarding
energy in $^{58}$Ni + Cu collisions \cite{eos1}. In this case
much larger mass shifts of $-75$ MeV/c$^2$ were observed in central
collisions, the shift became smaller with increasing impact parameter. That
the $(p, \pi^{\pm})$ pairs can also be successfully studied in central
collisions
of very heavy systems at $1$ AGeV bombarding energy was shown in \cite{trza94}.
This report contained a preliminary analysis of the data used in the present
publication. Although at $1$ AGeV energy the excitation of the $\Delta(1232)$
resonance is the dominating channel there is direct experimental evidence
from the identification of the $\eta$ meson \cite{berg94}, that at this energy
also higher-lying resonances, in particular the $N(1535)$ resonance, must
become excited. If this is the case then the resonance decay into the $2\pi$
channel also has to be present, and the analysis of $(p, \pi^{\pm})$
correlations will not yield a complete picture of the resonance distribution.
On the other hand, the $2\pi$ decay channel can be observed in the pion
transverse momentum spectra \cite{teis97a}.

The subsequent sections are organized as follows: The section 2
contains a short description of the experimental procedures. A more detailed
description can be found in \cite{fopi1}. The methods to extract the
invariant mass distributions of the resonances involved are presented in
section 3. These methods include the analyses of the measured $\pi^-$ and
$\pi^+$ transverse momentum spectra, and the analyses of the $(p, \pi^-)$
and $(p, \pi^+)$ correlations. In both cases the residual Coulomb
interaction between the charged pions and the positively charged baryon
distribution causes a perturbation \cite{fopi1,fopi2}, which is of
opposite sign for $\pi^-$ and $\pi^+$ and which we compensate by
analyzing the average of the $\pi^-$ and $\pi^+$ spectra, respectively by 
taking the average invariant mass distribution from $(p, \pi^-)$ and 
$(p, \pi^+)$. The justification to apply
this procedure is given in the appendix 6.1. The experimental results
are presented in section 4, section 5 contains their discussion and the
summary.

\section{Experimental procedures}
The reactions $^{197}$Au + $^{197}$Au and  $^{58}$Ni + $^{58}$Ni were
studied at nominal beam energies of $1.06$, respectively $1.06, 1.45$ and
$1.93$ AGeV. The $^{197}$Au and $^{58}$Ni beams were accelerated by the
$UNILAC/SIS$ accelerator combination of the $GSI$ /Darmstadt. The duty
cycle was 75\% with a spill length of $4$ s. The average beam intensities
varied between $1 \cdot 10^5$ and $5 \cdot 10^5$ particles per spill. The
targets consisted of self supporting foils with a thickness of
$100 \mu$m(Au), respectively $270 \mu$m(Ni). This corresponds to an
interaction probability of 0.5\%. The energy loss of the beams in the
targets is of around $0.01$ AGeV and was neglected.

The particles produced by the Au + Au and Ni + Ni reactions were
detected by the $FOPI$ detector which is a modular detection system with
almost $4\pi$ coverage. Of particular importance for the present investigation
is the central drift chamber $CDC$ which is mounted inside the superconducting
magnet with a solenoidal field of $0.6$ T strength. The $CDC$ has complete
cylindrical symmetry, it covers the polar angles $\vartheta$ from $32^{\circ}$
to $150^{\circ}$ in the laboratory frame and allows the particles masses and
charges to be simultaneously determined. For charged pions the geometrical
boundaries of the $CDC$ guarantees a detection efficiency of better than 65\%,
in particular the midrapidity region $|Y^{(0)}| < 0.1$ is fully covered above a
transverse momentum limit $p_{\rm t}^{(0)} > 0.65$ (for the definition of the
variables used here, see below). Notice, however, that to distinguish the
$\pi^+$ from protons an additional momentum limit $p < 0.65$ GeV/c was
introduced \cite{fopi1}. This limit which corresponds to $p_{\rm t}^{(0)} =
3.2$ at midrapidity, has no consequences for the situation
shown in Fig.1, and it is not required for the $\pi^-$. For protons
the geometrical $CDC$ boundaries $32^{\circ} < \vartheta < 150^{\circ}$ reduce
their detection efficiency to approximately 45\%.

The situation for particles produced in symmetric reactions at $1.93$ AGeV
bombarding energy is depicted in Fig.1. The fully drawn curves display the
lower and upper geometrical limits $\vartheta_{\rm l} = 32^{\circ}$ and
$\vartheta_{\rm u} = 150^{\circ}$ in the laboratory frame, the dotted curves
are drawn for pions respectively protons with $cm$ kinetic energies of $300$
MeV. The shaded area displays the momentum distribution of identified
$\Delta(1232)$ resonances, i.e. the decay proton and pion were measured by the
$CDC$. The resonances are Monte Carlo generated using a fixed mass $m_{\Delta}
= 1232$ MeV/c$^2$ and an isotropic
Boltzmann distribution with temperature $T = 173$ MeV, which corresponds to a
mean kinetic energy of $300$ MeV. Because of the decay into a
correlated $(p, \pi)$ pair the $\Delta(1232)$ resonance can be observed
in regions of the momentum space which are prohibited for single pions
or protons by the $\vartheta_{\rm l} = 32^{\circ}$ cut. At midrapidity the
$\Delta(1232)$ resonance is detected for transverse momenta $p_{\rm t}^{(0)}
> 0.5$. Notice that the symmetry law
\begin{eqnarray}
f(\Theta , \Phi) &=& f(\pi - \Theta , \pi + \Phi) \hspace{2mm},
\end{eqnarray}
which holds for all symmetric reactions guarantees that the region $Y^{(0)}
> 0$ can be reconstructed by reflection around midrapidity from the region
$Y^{(0)} < 0$. In this way the properties of around 50\% af all
$\Delta(1232)$ resonances are known. The remaining ones which cannot be
reconstructed
decay into a $(p, \pi)$ pair, where either one of the two particles or both
are not detected within the geometrical boundaries of the $CDC$. 
Under the assumptions made, i.e. 
isotropic resonance distribution and isotropic resonance decay, the efficiency 
of the $CDC$ for detecting resonances can be calculated as a function of the 
mass and the kinetic energy of the resonance.
\begin{figure}
\epsfxsize=8.5cm
\epsffile[0 150 565 440]{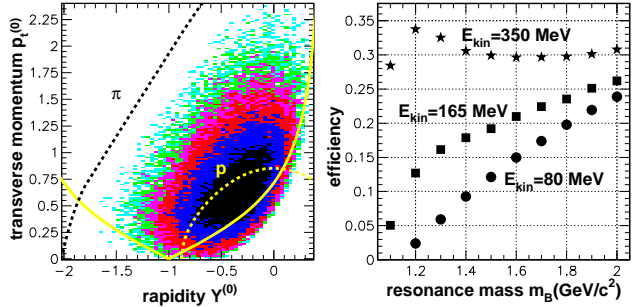}
\caption{Left panel: The reconstructed momentum distribution of thermal
$\Delta(1232)$ resonances with a temperature of 173 MeV. The full curves
display the geometrical limits of the $CDC$, the dashed curves display the
momenta of pions, respectively protons with a kinetic energy of 300 MeV.
Right panel: The efficiency of reconstructing resonances with masses
$m_{\rm B}$ and with kinetic energies $E_{\rm kin}$ from $(p, \pi)$ pairs
detected by the $CDC$}
\end{figure}
The efficiencies to detect resonances with masses $m_{\rm B}$ and
kinetic energies $E_{\rm kin}$  are plotted in the right panel of Fig.1, without
employing symmetrization. The detection efficiencies decrease rapidly with
mass for small kinetic energies. Without the proper efficiency corrections,
which in this case are solely due to the incomplete $4\pi$ acceptance of the
detector, the contributions from small mass values $m_{\rm B}$ would be
reduced. Notice that from the results published in \cite{eos2} for
the Au + Au reaction at $1$ AGeV bombarding energy, the expected mean
kinetic energy of the $\Delta(1232)$ resonance with mass $m_{\Delta} =
1232$ MeV/c$^2$ is $<E_{\rm kin}> = 200$ MeV. To obtain the efficiency
correction for a given mass the data shown in Fig.1 have to be weighted by
the energy distribution of that mass and integrated.

In addition to the $CDC$, the information from the forward plastic
scintillation wall $PLA$ mainly served trigger purposes. The $PLA$ is a
high-granularity component of the $FOPI$ detector and covers the polar
angles $\vartheta$ from $7^{\circ}$ to $30^{\circ}$ with full cylindrical
symmetry.
It separates particles only according to their charges. In the present analysis
the particle multiplicities $n_{\rm PLA}$ measured with the $PLA$ were required
to be larger than 80\% of the maximum multiplicity which is defined by
the lower boundary of the $PM5$ multiplicity bin, see \cite{fopi1,fopi2}. 
This multiplicity criterion selects near-central collisions
in the impact parameter range from $0$ to $0.3\,b_{\rm max}$, where
$b_{\rm max}$ is the maximum impact parameter of a given reaction. On the
average the fireball then contains around $3/4$ of
the maximum number $A_0 = A_{\rm proj} + A_{\rm targ}$ of nucleons.

In the present study we shall use the identical conventions which were used
in \cite{fopi1,fopi2}. All quantities which refer to the
laboratory($lab$) or target frames will be labeled by small letters. The
center of mass($cm$) or fireball frames are generally characterized
by capital letters. Exceptions are those
quantities, like the transverse momentum, which remain unchanged under the
transformation of frames. 
For the transverse momentum we sometimes use the quantity $p_{\rm t}^{(0)}$ 
where the index $^{(0)}$ indicates a normalization by the factor
$(A \cdot P_{\rm proj} /A_{\rm proj})^{-1}$
and where $A$ is the particle's mass number. Similarly the
rapidity Y in the $cm$ frame is normalized by the factor
$(Y_{\rm proj})^{-1}$
and this quantity is labeled $Y^{(0)}$. In the present experiments the
normalization factors for the pion transverse momenta have the values
$9.47, 8.09, 7.02$ (GeV/c)$^{-1}$ for $1.06, 1.45, 1.93$ AGeV bombarding
energy, to obtain the equivalent values for protons these numbers should be
multiplied with a factor $0.149$. For the rapidities the corresponding
normalization factors are $1.44, 1.26, 1.12$.

The data from the drift chamber $CDC$ were analyzed by using three
different tracking algorithms as described in \cite{fopi1}. Differences in
tracking are mainly observed in case of the Au + Au reaction, they consist
in a slightly different (10\%) efficiency with which negative pions are
identified. Equal tracking efficiencies can be obtained by adjusting the
limits in which the tracking parameters are allowed to vary \cite{fopi1}.
We have verified that our results do not depend on these limits, and agree
for the different versions of the tracking algorithm. Our final data, i.e.
transverse momentum and invariant mass spectra, are the average of three
independent analyses with these different trackers, and this data sample has
been used before in our study of the pion production in Au + Au \cite{fopi1}
and Ni + Ni \cite{fopi2} reactions. The accuracy with which the
particle momenta can be measured with the $CDC$ was determined in
\cite{fopi1}. For the transverse momentum the relative standard
deviation $\sigma(p_{\rm t}) / p_{\rm t}$ is $0.04$ for
$p_{\rm t} < 0.5$ GeV/c, and then increases to $0.1$ for transverse momenta
$p_{\rm t} = 1.5$ GeV/c. The accuracy in determining the azimuthal angle
$\varphi$ is better than $1^{\circ}$, the accuracy to determine the polar angle
$\vartheta$ is better for protons ($\sigma(\vartheta) = 3^{\circ}$) than for
pions ($\sigma(\vartheta) = 5^{\circ}$). These values may be used to estimate
the accuracy with which the mass $m_{\rm B}$ of a baryon resonance {\rm B}
can be determined. Without going into the details which will be more fully
presented in the next section, the accuracy estimate is
\begin{eqnarray}
&&\frac{\sigma(m_{\rm B})}{m_{\rm B}} =
\frac{m_{\rm B}^2 - m_0^2}{m_{\rm B}^2} \times \nonumber\\
&& 0.5 \sqrt{\left(\frac{\sigma(p_{\rm p})}{p_{\rm p}}\right)^2 +
\left(\frac{\sigma(p_{\pi})}{p_{\pi}}\right)^2 +
(\sigma(\alpha)\tan\alpha)^2} \hspace{2mm},
\end{eqnarray}
where $m_0$ is the sum of the proton and pion masses and $\alpha$ is the
angle between $\bf p_{\rm p}$ and $\bf p_{\pi}$. Taking representative
values, i.e. \\
\begin{center} \[
\hspace{2mm}
\frac{\sigma(p_{\rm p})}{p_{\rm p}} = 0.06 \hspace{2mm} , \hspace{2mm}
\frac{\sigma(p_{\pi})}{p_{\pi}} = 0.04 \hspace{2mm} , \hspace{2mm}
\sigma(\alpha)\tan\alpha = 0.09 \]
\end{center}
one obtains for, e.g. the mass of the $\Delta(1232)$ resonance an error of
$\sigma(m_{\Delta}) = 20$ MeV/c$^2$, with decreasing accuracy for larger masses
$m_{\rm B}$. This estimate is certainly sufficient to determine 
the resonance mass distribution, it is however not representative, since
a second source of systematic errors comes from the reconstruction of
the combinatorial background of uncorrelated $(p, \pi^{\pm})$ pairs, as will be
discussed in the next section. In the present analysis the number of directly
measured $(p, \pi^{\pm})$ pairs is larger than $10^6$ for a given reaction, to
reconstruct the background for this reaction we have used approximately $10$
times more pairs. The statistical errors are therefore not the decisive
factor in our analysis.

\section{Methods of analysis}
We have employed two independent procedures to determine the mass distribution
of baryon resonances excited in near-central collisions of identical nuclei.
In the first method the pion $p_{\rm t}$ spectrum is defolded to yield the
mass distribution, in the second method the invariant mass distribution
of the correlated $(p, \pi^{\pm})$ pairs is deduced from the data. 
Both methods rely on the
assumption that the resonances involved in the production of pions decay
via the emission of only one pion. This assumption is not necessarily
fulfilled as was pointed out in the introduction. If the two-pion decay is
present it effects the results of the two methods in different ways. In the
first case the presence of the two-pion channel will cause a distortion
of the mass spectrum, in the second case it will mainly contribute to the
background. It is therefore not expected that both methods should yield
identical results, however these results should be reasonably close since
the two-pion channel is expected to be weak \cite{teis97a}. We therefore also
take a reasonable agreement as an indication that systematic errors are
sufficiently small.
\begin{table*}
\caption{Temperatures $T$ and flow velocities $\beta$ used in defolding the
measured $p_{\rm t}$ spectra of $\pi^{\pm}$. The $r^{\pm}$ and
$r_{\rm e}^{\pm}$ values give the measured, respectively calculated (eq.(6))
fraction of correlated $(p, \pi^+)$ respectively $(p, \pi^-)$ pairs,
$\kappa_{\rm p}^{\pm}$ is the probability that the proton escapes
unscattered. The first row is for the Au + Au reaction, the next three
rows for the Ni + Ni reactions}
\begin{center}
\begin{tabular}{|c||c|c||c|c|c||c|c|c|}
\hline
 energy  &    $T$     &     $\beta$     & $r^+$ & $r_{\rm e}^+$ &
 $\kappa_{\rm p}^+$ & $r^-$ & $r_{\rm e}^-$ & $\kappa_{\rm p}^-$ \\
 (AGeV) & (MeV) &  & (\%) & (\%) &  & (\%) & (\%) &  \\
\hline
1.06 & 81 & 0.32 & 0.75 $\pm$ 0.25 & 0.75 &  1.0 $\pm$ 0.3  &  0.6 $\pm$ 0.2  &
0.22 & 2.7 $\pm$ 0.9  \\
\hline
1.06 & 79 & 0.23 &  1.0 $\pm$ 0.3  & 2.17 & 0.46 $\pm$ 0.15 & 0.75 $\pm$ 0.25 &
0.68 & 1.1 $\pm$ 0.35 \\
\hline
1.45 & 84 & 0.29 & 0.95 $\pm$ 0.3  & 2.16 & 0.44 $\pm$ 0.15 &  0.6 $\pm$ 0.2  &
0.71 & 0.85 $\pm$ 0.3 \\
\hline
1.93 & 92 & 0.32 & 1.05 $\pm$ 0.3  & 2.19 & 0.48 $\pm$ 0.15 &  0.6 $\pm$ 0.2  &
0.71 & 0.85 $\pm$ 0.3 \\
\hline
\end{tabular}
\end{center}
\end{table*}

\subsection{The analysis of the $p_{\rm t}$ spectra}
The defolding of the experimental $p_{\rm t}$ spectra of charged pions was
the earliest method applied by Brockmann et al. \cite{broc84,sand85} to
deduce the mass of the $\Delta(1232)$ resonance. Recently this technique
was also applied by \cite{muen97} to defold the $p_{\rm t}$ distribution of
$\pi^+$ from Au + Au reactions. It is well known that the $p_{\rm t}$
spectrum of $\pi^+$ differs from the $p_{\rm t}$ spectrum of $\pi^-$ for
momenta $p_{\rm t} < 0.3$ GeV/c, and that this difference is caused by the
Coulomb interaction between the pions and the nuclear matter distribution 
\cite{fopi1,fopi2}. In the appendix 6.1 we shall explore this effect more
closely and show that the Coulomb effects cancel when the $p_{\rm t}$ spectra
of $\pi^+$ and $\pi^-$ are combined. We follow this prescription in our
analysis and add the $p_{\rm t}$ spectra of $\pi^-$ and $\pi^+$, measured in
a narrow rapidity gap $|Y^{(0)}| < 0.1$. The individual $p_{\rm t}$ spectra
of $\pi^-$ and $\pi^+$ have almost identical cross sections for
$p_{\rm t} > 0.3$ GeV/c \cite{fopi2}, this part of the $p_{\rm t}$ spectra
thus supplied the required normalization factor. The pion
$p_{\rm t}$ spectrum is  defolded under the hypothesis that it is entirely
due to the one-pion decay of excited baryon resonances:
\begin{eqnarray}
\frac{d \sigma_{\rm j}}{d p_{\rm t}} =
\sum_{\rm i=1}^{\rm n} \frac{d \sigma_{\rm ij}(T, \beta)}{d p_{\rm t}} \cdot 
f(m_{\rm i}) \hspace{5mm} ; \hspace{5mm} max({\rm j}) \ge {\rm n} \; ,
\end{eqnarray}
where $d \sigma_{\rm ij} / d p_{\rm t}$ is the midrapidity transverse momentum
distribution of pions which are emitted in the decay of a baryon resonance
with mass $m_{\rm i} > m_0$. In \cite{soll91} the method is described to
calculate $d \sigma_{\rm ij} / d p_{\rm t}$.
This distribution depends only weakly on the
kinetic energy distribution of the resonances, a phenomenon which was first
noticed by Brockmann et al. \cite{broc84,sand85}. Using the formula of
Siemens and Rasmussen \cite{siem79} the kinetic energy distribution is usually
parametrized by the nuclear temperature $T$ and the flow velocity $\beta$.
For these we have used in our analysis the values published in \cite{eos2}
for the Au + Au, and in \cite{fopi5} for the Ni + Ni systems.
These values of $T$ and $\beta$ are listed in table 1. The defolding
of eq.(3) was performed with the $EM$ algorithm \cite{demp77}. The $EM$
algorithm finds after a sufficient number of iterations the
Maximum-Likelihood estimate of $f(m_{\rm i})$. Its performance and accuracy,
particularly when applied to defolding the pion $p_{\rm t}$ spectra, was
studied in detail in \cite{durs95}. It was shown that the result is unambiguous
and independent of the initial $f(m_{\rm i})$ distribution, when the functions
$d \sigma_{\rm ij}(T, \beta) / d p_{\rm t}$ are linear independent.
The mass resolution obtained by the defolding was 
studied as a function of the number of iterations by means of 
the point-spread-functions of $m_{\rm i}$. Allowing for infinite number of 
iterations the mass resolution reaches for all $m_{\rm i}$ its ideal value of 
10 MeV/c$^2$ in accordance with the width of the used mass bins. 
Limiting the number 
of iterations the mass resolution decreases especially for larger masses. The 
number of iterations is, however, bounded by the statistical errors of the 
measured $d \sigma_{\rm j} /d p_{\rm t}$ spectrum.
The iterations are stopped when the calculated spectrum
(right hand side of eq.(3)) becomes identical with the measured
$d \sigma_{\rm j} / d p_{\rm t}$ spectrum within these errors. Otherwise the
algorithm would run into describing the statistical fluctuations. The mass
resolution reached for masses in the important region $m_{\rm i} < 1400$
MeV/c$^2$ was 10 $-$ 30 MeV/c$^2$, for masses $m_{\rm i} < 1700$ MeV/c$^2$ it
still remained below 50 MeV/c$^2$. These errors are the statistical
uncertainties for a mass bin $m_{\rm i}$, the accuracy to determine the
mean value of the mass distribution $f(m_{\rm i})$ is much better and
in fact mainly determined by systematic errors which we have deduced from
a comparison between our two different techniques to obtain $f(m_{\rm i})$.

\subsection{The analysis of the $(p, \pi^{\pm})$ pairs}
The invariant mass of a $(p, \pi^{\pm})$ pair can in principle be calculated
from the measured momenta ${\bf p}_{\rm p}$ and ${\bf p}_{\pi}$ via
\begin{eqnarray}
m_{\rm B} = \sqrt{(e_{\rm p} + e_{\pi})^2 - ({\bf p}_{\rm p}
+ {\bf p}_{\pi})^2} \hspace{5mm} \mbox{where} \hspace{1cm} \\
e_{\rm p} = \sqrt{(p_{\rm p})^2 + (m_{\rm p})^2} \hspace{5mm} \mbox{and}
\hspace{5mm} e_{\pi} = \sqrt{(p_{\pi})^2 + (m_{\pi})^2}. \nonumber
\end{eqnarray}
For many of such pairs the resulting mass distribution
$d n^{\pm}_{({\rm meas})} / d m_{\rm B}$ contains a large combinatorial
background contribution $d n^{\pm}_{({\rm back})} / d m_{\rm B}$ from
uncorrelated \linebreak
$(p, \pi^{\pm})$ pairs, which has to be subtracted from the
former in order to obtain the required mass distribution
$d n^{\pm} / d m_{\rm B}$ of truly correlated pairs:
\begin{eqnarray}
\frac{d n^{\pm}}{d m_{\rm B}} = \frac{d n^{\pm}_{({\rm meas})}}{d m_{\rm B}} -
\left(1 - r^{\pm}\right) \frac{d n^{\pm}_{({\rm back})}}{d m_{\rm B}}
\hspace{2mm} .
\end{eqnarray}
In this expression $d n^{\pm}_{({\rm meas})} / d m_{\rm B}$ and
$d n^{\pm}_{({\rm back})} / d m_{\rm B}$ are normalized to equal intensity,
and $r^{\pm}$ is the ratio of the number of "true" pairs to the number of
"random" pairs. It is assumed that $r^{\pm}$ is constant, i.e. does not depend
itself on $m_{\rm B}$, which follows from the condition that
$d n^{\pm}_{({\rm back})} / d m_{\rm B}$ is correctly known. The analysis
consists of determining the size of $r^{\pm}$ and the shape of the
background contribution $d n^{\pm}_{({\rm back})} / d m_{\rm B}$.

The generally accepted way to obtain the background spectrum is based on
the method of event mixing. This method takes the proton and the pion of
a pair not from the same event but from different events, using the fact
that all "true" correlations do not exist in mixed pairs, but that "random"
correlations will survive the mixing. Besides the true correlations due to
the resonance decay there are additional correlations due to the reaction
dynamics as, for example, the particle focussing with respect to the event
plane. These latter correlations are also destroyed by event mixing and
therefore are not removed when subtracting the mixed spectrum from the
measured spectrum unless special precautions are taken. These require that
for the pairs measured in a certain event the corresponding mixed pairs
come from events with identical proton and pion multiplicities and identical
orientation of the event plane. The latter requirement can always be
fulfilled by the proper rotation of both events around the beam axis. The
condition of equal multiplicities in the measured and mixed spectra also
ensures that correlations due to pion rescattering in spectator matter
\cite{fopi1,fopi2} are properly subtracted.
\begin{figure}
\epsfxsize=8.5cm
\epsffile[15 140 565 440]{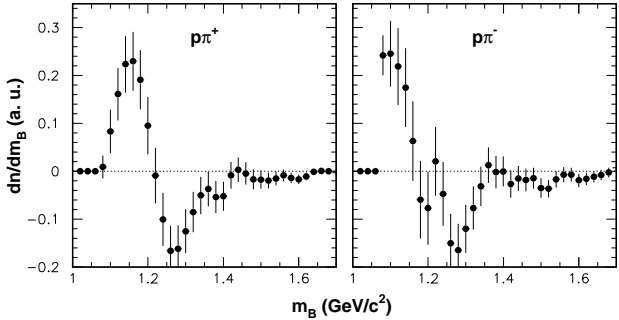}
\caption{The signal for the existence of correlated $(p, \pi^+)$ (left panel), 
respectively $(p, \pi^-)$ pairs (right panel) in the reaction Au + Au at 
1.06 AGeV bombarding energy. The total intensity in the correlation 
spectra is normalized to zero}
\end{figure}

The general question of how reliably the
uncorrelated background spectrum can be reconstructed by means of the event
mixing technique was studied in detail
by L'Hote \cite{hote94}. In the majority of cases the difference between
the measured and mixed spectrum (right hand side of eq.(5)) includes, besides
the invariant mass distribution from true pairs, a residual contribution from
random pairs which becomes the smaller, the smaller the differences between
the phase space distributions of protons and baryon resonances. We have
confirmed these conclusions with our Monte Carlo studies, and have found that
in the present case, where the decay is dominated by the $\Delta(1232)$
resonance and where the temperatures $T$ are moderate, the contribution of
random pairs to the right hand side of eq.(5) is small. In addition we take
the agreement with the results from the defolding procedure as independent
evidence that the event mixing technique is reliable in the present analysis.

With the method at hand to determine $d n^{\pm}_{({\rm back})} / d m_{\rm B}$
it is easy to prove that the measured $d n^{\pm}_{({\rm meas})} / d m_{\rm B}$
spectrum contains a contribution from correlated $(p, \pi^{\pm})$ pairs.
For this purpose we assume $r^{\pm} = 0$ and
calculate $d n^{\pm} / d m_{\rm B}$ according to eq.(5). The
resulting spectrum has total intensity zero, the correlations appear as
regions of positive respectively negative intensities for certain values of
$m_{\rm B}$. This is shown in Fig.2 for the Au + Au reaction
at $1.06$ AGeV bombarding energy. 
The correlation signals are strikingly strong and they
provide the undisputable evidence that part of the $(p, \pi^{\pm})$ pairs are
correlated due to the decay of baryon resonances.

In order to extract the
mass distribution of these resonances the ratios $r^{\pm}$ have to be
experimentally determined. The determination is done in principle, i.e. when
statistical fluctuations are negligible, by the requirement that the right hand
side of eq.(5) should be positive. In reality these fluctuations cause part
of the $d n^{\pm} / d m_{\rm B}$ spectrum to attain negative values.
From Monte Carlo calculations with similar counting statistics as in the
experiments we have concluded that the negative part may vary between
2\% and 5\% of the total positive intensity. These fractions provided the
proper criteria for the background subtraction and the determination of the
$r^{\pm}$ values within an uncertainty of 30\%. Furthermore, the resulting
$d n^{\pm} / d m_{\rm B}$ distribution depends only weakly on the chosen
value of $r^{\pm}$ in the region where $d n^{\pm} / d m_{\rm B}$ has its
maximum. The values obtained for $r^+$ respectively $r^-$ are listed
in table 1.

The relative contribution of correlated pairs $r^{\pm}$ can
be calculated from the measured proton multiplicity $n_{\rm p}$ and the
known detection efficiency for protons $\epsilon_{\rm p}$:
\begin{eqnarray}
r^{\pm} = \frac{\epsilon_{\rm p}}{n_{\rm p}} \alpha_{\rm I}^{\pm}
\kappa_{\rm p}^{\pm} \xi_{\pi} \hspace{2mm},
\end{eqnarray}
where $\alpha_{\rm I}^{\pm}$ is an isospin dependent factor which
takes into account that part of the detected pions are correlated with
undetected neutrons,
$\kappa_{\rm p}^{\pm}$ is the
probability that the proton reaches the detector without another
proton - nucleon interaction, and $\xi_{\pi}$ describes the fraction of the
pions which are correlated, i.e. due to resonance decay.
The factor $\alpha_{\rm I}^{\pm}$ depends on the
isospin of the decaying resonance and is different for $\pi^-$ and $\pi^+$.
For isospin $I = 1 / 2$ one obtains $\alpha_{1/2}^- = 1$ and
$\alpha_{1/2}^+ = 0$ for $\pi^-$ and $\pi^+$, respectively, for $I =
3 / 2$ the corresponding factors are $\alpha_{3/2}^- = (N + 2 \cdot Z) /
(10 \cdot N + 2 \cdot Z)$ and $\alpha_{3/2}^+ = 9 \cdot Z / (10 \cdot Z +
2 \cdot N)$. Notice that $I = 1 / 2$ resonances cannot be observed in the
$(p , \pi^+)$ channel. The isospin factors for $I = 3 / 2$ resonances depend
on the $N$ over $Z$ ratio. For the Au + Au reaction one obtains
$\alpha_{3/2}^+ = 0.69, \alpha_{3/2}^- = 0.21$, for the Ni + Ni reactions
these factors are $\alpha_{3/2}^+ = 0.74, \alpha_{3/2}^- = 0.24$.
Assuming a negligible contribution from thermal pions ($\xi_{\pi} = 1$)
and that only the decay of
$I = 3 / 2$ resonances contribute to the yield of correlated pairs
($\alpha_{\rm I}^{\pm} = \alpha_{3/2}^{\pm}$), eq.(5) provides an upper limit
$r_{\rm e}^{\pm}$ of $r^{\pm}$ when $\kappa_{\rm p}^{\pm} = 1$. The values
of $r_{\rm e}^{\pm}$ are quoted in table 1. The direct comparison between
$r^{\pm}$ and $r_{\rm e}^{\pm}$ gives the experimental values of
$\kappa_{\rm p}^{\pm}$ also quoted in table 2. The errors of
$\kappa_{\rm p}^{\pm}$ orginate from the uncertainty of $r^{\pm}$ ($30\%$).
In the next section we shall discuss the consequences which follow from
these $\kappa_{\rm p}^{\pm}$ values.

\begin{figure}
\epsfxsize=8.5cm
\epsffile[70 115 470 475]{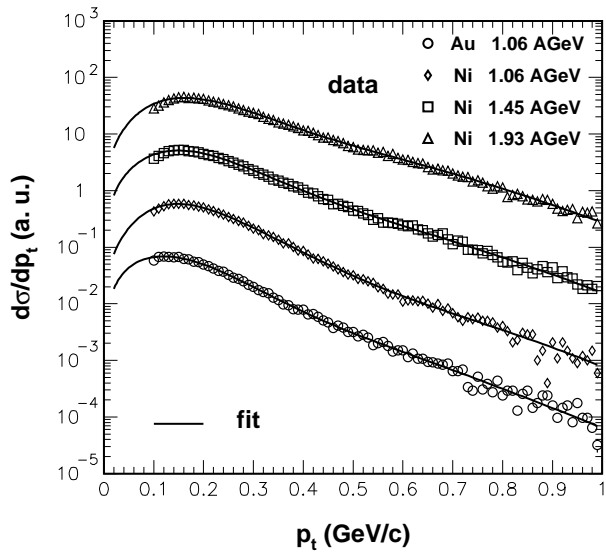}
\caption{The measured (open symbols) and calculated (full curves) transverse
momentum spectra in the rapidity range $|Y^{(0)}| < 0.1$ for the reactions
Au + Au (Au) and Ni + Ni (Ni) at the stated incident energies}
\end{figure}

\section{Experimental results}
\subsection{Results from the $p_{\rm t}$ distributions}
The pion $p_{\rm t}$ spectra obtained by adding the measured $\pi^+$ and
$\pi^-$ spectra in the rapidity range $-0.1 \le Y^{(0)} \le 0.1$, are
displayed in Fig.3. For the defolding by means of the method presented in
subsection 3.1 we have used the temperatures $T$ and the radial flow
velocities $\beta$ determined by \cite{eos2} and \cite{fopi5} and listed in
table 1.
The mass distributions are shown as histograms in Fig.4. The shaded
areas represent the uncertainties of the defolding technique. These
uncertainties are firstly due to the possible changes in the values of $T$
and $\beta$, and secondly due to the possible contribution of the $2 \pi$
decay channel to the low-momentum part of Fig.3. Considering the first point
first it should be mentioned that a recent analysis \cite{fopi6} of the
reaction Au + Au
at 1.06 AGeV incident energy indicates a lower value of the temperature
($T \approx$ 50 MeV) and a larger value for the collective velocity
($\beta \approx$ 0.5) yielding a 10\% larger mean kinetic energy of the
$\Delta(1232)$ resonance than deduced from table 1. In order to estimate
the changes which may result from different values of the temperature
and the flow we have repeated the defolding of all pion $p_{\rm t}$ spectra
using a 40\% smaller value of $T$ and values
of $\beta$ which correspond to a 10\% increase in mean kinetic
energy. The uncertainties of $f(m_{\rm i})$ are given by the shaded areas
in Fig.4. The uncertainty contributions from the decay into two pions were
estimated by starting the defolding of the $p_{\rm t}$ spectra first at
momentum values $p_{\rm t} = 0.1$ GeV/c and then at $p_{\rm t} = 0.15$ GeV/c.
Also these changes are included in the shaded areas in Fig.4.

The main features of the mass distributions obtained by the defolding
technique are their shifts of approximately $-60$ MeV/c$^2$ away from the mass
of the free $\Delta(1232)$ resonance, shown by the arrows in Fig.4,
and their weak extensions towards
larger masses. These tails of the mass distributions are observed for all
reactions, but it appears to be strongest in case of the Ni + Ni reaction at
$1.93$ AGeV bombarding energy.

\subsection{Results from the $(p, \pi^{\pm})$ pairs}
\begin{figure}
\epsfxsize=8.5cm
\epsffile[10 30 550 550]{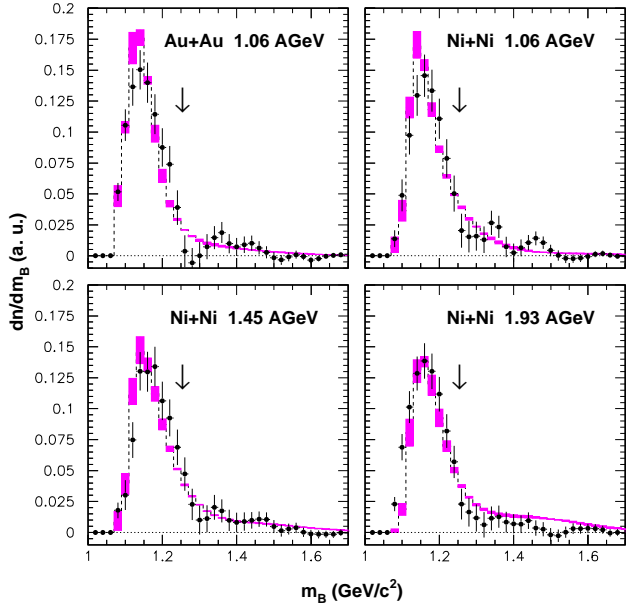}
\caption{The invariant mass spectrum of baryon resonances excited by the
reactions stated in each picture. The shaded areas correspond to
the analysis of the measured transverse momentum spectra of $\pi^{\pm}$, the 
full points to the analysis of the measured $(p, \pi^{\pm})$ pairs. The arrows
point to the maximum of the free $\Delta(1232)$ mass distribution}
\end{figure}
We first consider the
effects which the residual Coulomb interaction between charged pions and the
nuclear matter distribution might exert onto the reconstruction of
invariant masses. In Fig.5
the invariant mass distributions obtained from $(p, \pi^+)$ respectively
$(p, \pi^-)$ pairs are shown for all reactions. These distributions are
normalized to equal integrated intensity to allow for an easy comparison
between the $\pi^+$ and $\pi^-$ data. The values of $r^{\pm}$ deduced from this
analysis and the corresponding values of $\kappa_{\rm p}^{\pm}$ are listed in
table 1. In all reactions the values of $\kappa_{\rm p}^-$ are larger than
those of $\kappa_{\rm p}^+$. This is the behaviour one would expect if the
resonance mass distribution contains an $I = 1/2$ component with a mean free
mass above the $\Delta(1232)$ mass. The large value of $\kappa_{\rm p}^-$
in case of the Au + Au reaction, c.f. table 1, requires a 25\% contribution
of $I = 1/2$ resonances to reduce it to its regular size of $\kappa_{\rm p}^-
= \kappa_{\rm p}^+ \approx 1$. In case of the Ni + Ni reactions the
$\kappa_{\rm p}^{\pm}$ are of size $\approx$ 0.6 with a 20\% contribution of
$I = 1/2$ resonances. These apparent differences between the Au + Au and
Ni + Ni reactions are not understood yet. On the other hand, in all reactions
the maxima of the mass distributions shown in Fig.5 are shifted towards masses
below the $\Delta(1232)$ mass. In order to quantify these results without
taking resort to the Lorentz shape of a free resonance, we have calculated the
mean mass $m_{\Delta} = <m_{\rm B}>$ and the standard deviation
$\sigma_{\Delta} = \sqrt{ <m_{\rm B}^2> - <m_{\rm B}>^2}$
in the region of the prominent peak with an upper limit $m_{\rm B} < m_<$.
For $m_<$ we have chosen a value $m_<$ = 1300 MeV/c$^2$ which is motivated
by a recent theoretical calculation of the $\Delta(1232)$ mass shape
\cite{wein96} and which shall be further discussed in the next section, c.f.
Fig.8. The results for $m_{\Delta}$ and $\sigma_{\Delta}$
are quoted in the first two rows of each reaction entry in table 2. One notices
that $m_{\Delta}$ from $(p,\pi^+)$ pairs is on the average slightly larger
than from $(p,\pi^-)$ pairs, and that the opposite holds for $\sigma_{\Delta}$.
We interpret the first result as caused by the differences in the Coulomb
interactions between $\pi^+$ respectively $\pi^-$ and the baryons, whereas the
second result again indicates the existence of $I = 1/2$ contributions which
can only be present in the $(p, \pi^-)$ channel.
\begin{figure}
\epsfxsize=8.5cm
\epsffile[10 30 550 550]{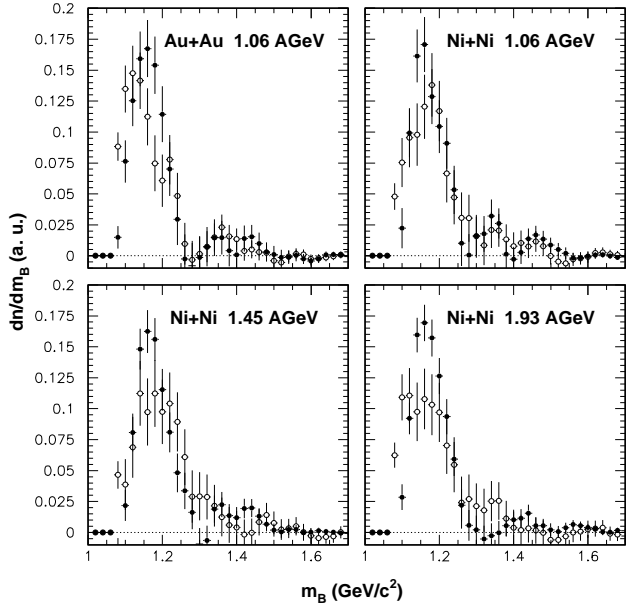}
\caption{The invariant mass spectra of correlated $(p, \pi^-)$ (open points) 
and $(p, \pi^+)$ (full points) pairs for the reactions stated in each picture}
\end{figure}

The influence of the Coulomb interaction is studied in more detail in the
appendix 6.1, where it is shown that similar to the prescription employed in
case of the defolding method, also in case of the $(p, \pi^{\pm})$
correlations the
influence can be minimized by adding the invariant mass spectra of
$(p, \pi^+)$ and $(p, \pi^-)$ pairs to yield the mass spectrum of
$(p, \pi^{\pm})$ pairs. On the latter spectrum we will base our further
analysis. 

The mass spectra from $(p, \pi^{\pm})$ pairs are shown, together with the
corresponding spectra from the defolding method, as points in Fig.4. For each
reaction these mass spectra obtained by two different techniques are very
similar in the range around the prominent $\Delta(1232)$ peak. Notice that
the efficiency corrections represented by Fig.1 were not applied to the
$(p, \pi^{\pm})$ data, but other corrections discussed in the appendix 6.2 were
applied. The reason that the influence of the detector geometry onto the
invariant mass does not need to be considered is due to our finding from
Monte Carlo simulations, that the corresponding decrease in the acceptance of
small masses is counterbalanced by an enhancement caused by the finite
detector resolution, see the appendix. Both effects are of similar size and
therefore cancel within the achieved experimental accuracies.

\subsection{The $\Delta(1232)$ mass shift and higher-lying resonances}
For a detailed analysis of the results presented in Fig.4 we have applied
the same methods used in the analysis of Fig.5. The mean mass $m_{\Delta}$
and the standard deviation $\sigma_{\Delta}$ were deduced in the region
$m_0 < m_{\rm B} < m_<$, their values are quoted in the last two rows of each
reaction entry in table 2. The mass shift of $\Delta m_{\rm B} = - 71 \pm 6$
MeV/c$^2$ found for the Au + Au reaction appears to be significantly larger
than the shifts found for the Ni + Ni reactions, which do not seem to depend
on incident energy and which amount to $\Delta m_{\rm B} = - 49 \pm 4$
MeV/c$^2$. The interpretation of these shifts is postponed to section 5.

Finally the Fig.4 illustrates that the measured mass distributions extend with
weak tails to masses above $m_{\rm B} > m_<$. Only the $(p, \pi^{\pm})$
data prove that these tails are associated with the excitation of resonances
above the $\Delta(1232)$ resonance. The defolding technique only proves that 
this conclusion is not in contradiction with the measured pion $p_{\rm t}$ 
spectra which could also contain thermal pions. The proof for the
excitation of higher resonances requires that
pions and protons from the decay of these resonances are correlated.
In order to confirm that pions with large momenta are indeed emitted by
these high resonances and that these pions are correlated with protons we
have reanalyzed the $(p, \pi^{\pm})$ data with a lower threshold of $p_{\rm t}
> 0.3$ GeV/c for the pion momentum. The invariant mass spectra, separated
according to the pion charges, are shown in Fig.6 for the Ni + Ni reaction
at 1.93 AGeV incident energy. As shaded areas the Fig.6 also displays the
invariant mass spectra obtained without the threshold on the pion momenta. 
Two conclusions can be derived from Fig.6:
\begin{itemize}
\item{The tail of the mass distribution is 
strongly associated with high $p_{\rm t}$ pions, since the applied $p_{\rm t}$
cut reduces only the yield from the mass region $m_{\rm B} < m_<$.
This is, of course, required by the decay kinematics of the resonances,
provided such resonances were excited.}
\item{The shape of the deduced mass distribution is different for $(p, \pi^+)$
and $(p, \pi^-)$ pairs, and the number of correlated pairs has increased
more for the $\pi^-$ than the $\pi^+$. This asymmetry again confirms our
earlier conclusions about the contribution of $I = 1/2$ resonances, since 
these cannot decay into $(p, \pi^+)$ pairs.}
\end{itemize}
\begin{figure}
\epsfxsize=8.5cm
\epsffile[20 120 550 400]{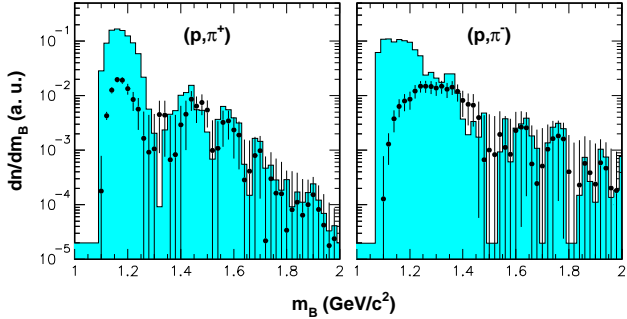}
\caption{The shaded areas show a logarithmic plot of the invariant mass 
spectra of correlated $(p, \pi^+)$ (left panel) and $(p, \pi^-)$ pairs 
(right panel) from the reaction Ni + Ni at 1.93 AGeV bombarding energy. 
The points display the equivalent spectra selecting pions with
$p_{\rm t} > 0.3$ GeV/c}
\end{figure}

\section{Summary and discussion}
In the present work we have shown that pions are produced in relativistic
heavy ion reactions by the decay of excited baryon resonances, with the
dominant contribution from the $\Delta(1232)$ resonance. The heavy ion
reactions studied were Au + Au at $1.06$ AGeV and Ni + Ni at $1.06$, 
$1.45$ and $1.93$ AGeV incident energies. The mass distributions of the baryon
resonances were deduced by means of two independent methods: Firstly by
defolding the pion $p_{\rm t}$ spectra measured at midrapidity by means of
eq.(3), and secondly by calculating the invariant masses of correlated 
$(p, \pi^+)$ respectively $(p, \pi^-)$ pairs by means of eq.(5). In both
analyses the Coulomb distortions due to the nuclear matter distribution were
compensated by adding the data obtained for $\pi^+$ and $\pi^-$. In case
of the $(p, \pi^{\pm})$ pairs this compensation was performed with the
invariant mass spectra which were obtained separately for $\pi^+$ and $\pi^-$.
The comparison between both suggests the additional excitation of $I = 1/2$
resonances, and it proves the persistence of the residual Coulomb distortion
in the $(p, \pi^-)$ respectively $(p, \pi^+)$ correlations.
\begin{table}[t]
\caption{The first 2 rows of each reaction entry give the values of 
$m_{\Delta}$ and $\sigma_{\Delta}$ deduced from $(p, \pi^+)$
respectively $(p, \pi^-)$ pairs, the third row gives the same for the
average  $(p, \pi^{\pm})$. The last row represents the results of the
$p_{\rm t}$ analysis including, in addition to $m_{\Delta}$ and
$\sigma_{\Delta}$, also the fraction $r_{{\rm B} > \Delta}$ of resonances with
masses above the $\Delta(1232)$ resonance, and the relative fraction 
$n_{\Delta}$ of baryons excited to the $\Delta(1232)$ resonance}
\begin{center}
\begin{tabular}{|c|c|c|c|c|}
\hline
 reaction  & $m_{\Delta}$  & $\sigma_{\Delta}$ & $r_{{\rm B} > \Delta}$ &
           $n_{\Delta}$ \\
           &  (MeV/c$^2$)  &    (MeV/c$^2$)    &                       &
               (\%)     \\
\hline
Au + Au   & 1160 $\pm$ 10 &    38 $\pm$ 5   &     &    \\
   at     & 1149 $\pm$ 10 &    48 $\pm$ 5   &     &    \\
1.06 AGeV & 1154 $\pm$ 10 &    43 $\pm$ 5   &     &    \\
          & 1154 $\pm$ 5  &    48 $\pm$ 3   & 0.08 $\pm$ 0.03 & 6.5 $\pm$ 2 \\
\hline
Ni + Ni   & 1173 $\pm$ 10 &    41 $\pm$ 5   &     &    \\
   at     & 1171 $\pm$ 10 &    53 $\pm$ 5   &     &    \\
1.06 AGeV & 1173 $\pm$ 10 &    47 $\pm$ 5   &     &    \\
          & 1175 $\pm$ 5  &    50 $\pm$ 3   & 0.10 $\pm$ 0.03 & 11.5 $\pm$ 3 \\
\hline
Ni + Ni   & 1177 $\pm$ 10 &    41 $\pm$ 5   &     &    \\
   at     & 1185 $\pm$ 10 &    56 $\pm$ 5   &     &    \\
1.45 AGeV & 1181 $\pm$ 10 &    48 $\pm$ 5   &     &    \\
          & 1176 $\pm$ 5  &    51 $\pm$ 3   & 0.15 $\pm$ 0.05 & 16 $\pm$ 4 \\
\hline
Ni + Ni   & 1174 $\pm$ 10 &    40 $\pm$ 5   &     &    \\
   at     & 1166 $\pm$ 10 &    56 $\pm$ 5   &     &    \\
1.93 AGeV & 1171 $\pm$ 10 &    48 $\pm$ 5   &     &    \\
          & 1182 $\pm$ 5  &    49 $\pm$ 3   & 0.21 $\pm$ 0.10 & 22 $\pm$ 6 \\
\hline
\end{tabular}
\end{center}
\end{table}

The results from both methods after the removal of the Coulomb distortions
were shown in Fig.4 and table 2. The agreement between the mass distributions
from the two independent methods is the prime indicator that systematic errors
in our analysis are small. Relative to the mass of the free $\Delta(1232)$
resonance the maxima of the deduced mass distributions are shifted towards
smaller masses. The mass shift is on the average $\Delta m_{\Delta} = -60 \pm
10$ MeV/c$^2$, where the lower limit corresponds to Au + Au, the upper limit
to Ni + Ni near-central reactions. The width of the $\Delta(1232)$ 
remains almost unchanged. The mass shift is more
than $2$ times larger than found in the preliminary analysis of \cite{trza94}.
The main reason for the deviation is that $(p, \pi^{\pm})$ pairs with small
relative angles between the particle momenta are not excluded in the present
analysis as they were in \cite{trza94}, see also the discussion in the
appendix 6.2. Our analysis of the $(p, \pi^{\pm})$ pairs suggests
that at least 50\% of the detected pions are correlated with protons, i.e.
they originate from the decay of baryon resonances. For the remaining part,
the correlations are either destroyed by the residual scattering of the proton
by nucleons, or those pions are of direct origin. The agreement between the
mass spectra from the defolding and correlation analyses makes the former
possibility more likely. In particular the application of $p_{\rm t}$ cuts in
the correlation analysis indicates that firstly higher resonances then the
$\Delta(1232)$ resonance are involved in the pion production, and that secondly
at least part of these resonances have isospin $I = 1/2$.

The mass shift deduced presently and the almost unchanged width of the 
mass distribution are in close agreement with the recent results determined 
from $(p, \pi^+)$ pairs in central Ni + Cu collisions at $1.97$ AGeV 
bombarding energy in \cite{eos1}. Nevertheless it should be pointed out that  
the resonance formula of \cite{cugn88} used in the analysis of \cite{eos1} 
to describe the $\Delta(1232)$ mass distribution produces a tail which
overestimates our measured $p_{\rm t}$ spectrum as demonstrated in the upper
part of Fig.7. Notice that the inclusion of even higher-lying resonances would
only enhance the disagreement at large transverse momenta. We thus conclude 
that under the present conditions the mass distribution assigned to the
dominant $\Delta(1232)$ resonance cannot be adequately described by the
parametrization of \cite{cugn88}.

There are two conceivable causes for these
modifications of the resonance shape: The resonance masses are shifted
because of their nuclear environment or/and the resonances are in thermal
equilibrium with the hadronic matter at a low temperature.

The effect of the hadronic environment on the masses of hadrons was
theoretically studied in many publications, for a recent review c.f.
\cite{ko96}. The environment causes
in general a mass shift which can be either positive or negative and depends
on the hadronic density. The verification of this prediction is at the moment
of considerable interest, at $GSI$ energies the experiments at present mainly
focus on the
kaon mass \cite{kaos1,fopi4}. The situation with respect
to baryons appears less understood. In a recent study \cite{wein96} the mass
of the $\Delta(1232)$ resonance was calculated to be shifted by
$\Delta m_{\Delta} \approx  -10$ MeV/c$^2$ when corrections due to the nucleon
interaction in the $\pi$N loop of the $\Delta$ self energy are taken into
account. The weight function $f_{\Delta}(m_{\rm B})$ which replaces in this
calculation the free mass distribution of the $\Delta(1232)$ resonance is
used in our further analysis.
\begin{figure}
\epsfxsize=8.5cm
\epsffile[130 150 460 440]{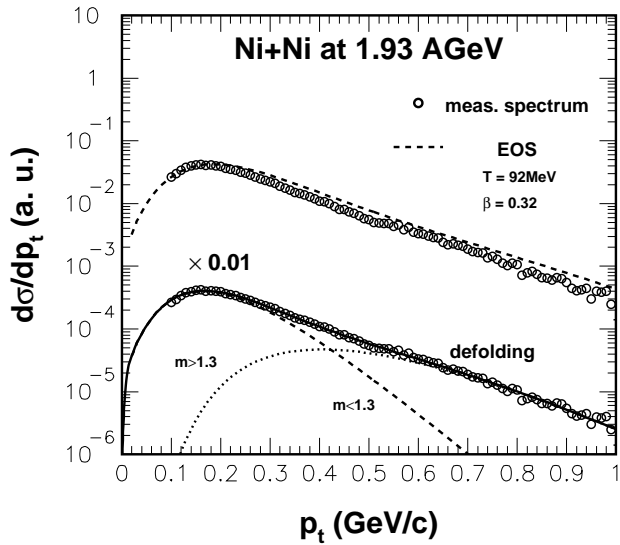}
\caption{Upper part: The $p_{\rm t}$ spectrum of $\pi^{\pm}$ measured(circles)
and calculated(dotted curves) using the $f_{\Delta}(m_{\rm B})$ distribution of
\protect\cite{eos1} and the $T$, $\beta$ values published in 
\protect\cite{fopi5}. Lower part: The decomposition of the $p_{\rm t}$
spectrum into the $\Delta(1232)$ component $(m_{\rm B} < 1.3)$ and
the higher resonances component $(m_{\rm B} > 1.3)$ }
\end{figure}

On the other hand a hadronic environment at low temperature $T$ and in
thermal equilibrium with the baryon resonances will result in an apparent mass
shift when these resonances decay.
Of course, the hypothesis that all hadrons are in thermal equilibrium needs
verification. But even if equilibrium is not established the development of
nuclear radial flow will cause the resonances to be populated predominantly at
the low-mass side in late stages of the reaction \cite{bass94}. In the
following we will assume for reasons of simplicity that the apparent mass
shift of the $\Delta(1232)$ resonance can be described by a thermal weight
function of Boltzmann type with temperature $T_{\Delta}$. 
An estimate of
$T_{\Delta}$, which henceforth shall be called temperature although it should
be considered primarily as fit parameter, is obtained by fitting the mass
spectra in Fig.4 by $f_{\Delta}(m_{\rm B}) \cdot exp(-(m_{\rm B}-m_0) /
T_{\Delta})$ in the mass region $m_{\rm B} < m_<$, where
$f_{\Delta}(m_{\rm B})$ was taken from the calculations of \cite{wein96}. 
The mass distributions obtained by both methods enter the fits with equal 
weights. The mass range 1150 MeV/c$^2$ $< m_{\rm B} <$ 1250 MeV/c$^2$ is 
however considered with a larger statistical weight, since the very low masses 
$m_{\rm B} < 1150$ MeV/c$^2$ are expected to be stronger affected by the 
possible systematic errors of both analysis methods, and the larger masses 
$m_{\rm B} > 1250$ MeV/c$^2$ are stronger affected by the contribution from 
higher resonances.
As an example the left panel of Fig.8 displays the fit to the invariant mass
distribution in case of the Ni + Ni reaction at 1.93 AGeV bombarding energy.
The required temperature is in this case $T_{\Delta} = 59 \pm 8$ MeV, for the
other reactions the temperatures are listed in table 3. 
The errors represent the uncertainties of the fits deduced by slightly 
shifting the fit region. 
Notice that the deduced 
values of $T_{\Delta}$ depend on the used parametrization of 
the mass distribution. The parametrization of \cite{cugn88} leads to values 
for $T_{\Delta}$ smaller by approximately 30\% than the ones listed in 
table 3.
\begin{table}
\caption{Temperatures $T_{\Delta}$ and radial flow velocities $\beta_{\Delta}$
deduced from a fit to the mass distributions in the range $m_{\rm B} <
1300$ MeV/c$^2$, and a fit to the kinetic energy distributions, c.f. Fig.8.
The first row is for the Au + Au reaction, the next three rows for the
Ni + Ni reactions}
\begin{center}
\begin{tabular}{|c|c|c|}
\hline
 energy(AGeV)  & $T_{\Delta}$(MeV) & $\beta_{\Delta}$ \\
\hline
      1.06     &   41 $\pm$ 5     & 0.46 $\pm$ 0.03  \\
\hline
      1.06     &   53 $\pm$ 8     & 0.39 $\pm$ 0.05  \\
\hline
      1.45     &   60 $\pm$ 8     & 0.47 $\pm$ 0.04  \\
\hline
      1.93     &   59 $\pm$ 8     & 0.53 $\pm$ 0.04  \\
\hline
\end{tabular}
\end{center}
\end{table}

The temperatures $T_{\Delta}$ obtained in this way are of the order of 50 MeV
and therefore much smaller than the 
temperatures deduced for the Au + Au and Ni + Ni reactions by other means
\cite{eos2,fopi5} and listed in table 1. As pointed out earlier in
\cite{fopi1}, these small values of $T_{\Delta}$ are linked to the 
enhancement of the pion spectrum at very small transverse momenta, which 
cannot be explained by the mass distribution of the {\bf free} $\Delta(1232)$
resonance, but requires the effective mass to be strongly shifted towards the
threshold mass by a medium with low temperature.

We interprete the calculated mass distribution of Fig.8 as the sole
contribution from the $\Delta(1232)$ resonance, which for $m_{\rm B} =
m_< = 1300$ MeV/c$^2$ practically has reached zero intensity.
The remaining yield not
accounted for is, because of its correlation with protons, at least partly
caused by higher lying resonances. As pointed out before this interpretation
is also suggested by the fact that the $(p,\pi^-)$ and
$(p,\pi^+)$ invariant mass spectra are different for large pion momenta and
that both pairs have different $\kappa_{\rm p}^{\pm}$ values, see table 1.
The upper estimate of the ratio between the contribution from high-lying
resonances and the one from the $\Delta(1232)$ resonance
is quoted in table 2 as $r_{\rm B > \Delta}$. These values
were extracted from the results of the defolding technique, since the analysis
of the correlation results, also shown in Fig.4, requires information about the
admixtures of $I = 1/2$ and $I = 3/2$ states in the tails of the invariant mass
distributions which does not exist with the required accuracy. We have used
the latter results, however, to obtain the quoted errors of $r_{\rm B >
\Delta}$. The contribution of the higher lying resonances to the $p_{\rm t}$
spectrum, obtained in this way, is shown in the lower part of Fig.7. It
demonstrates the difference to the alternative where the high-momentum tail
is interpretated as mainly due to thermal pions: The distribution from
thermal pions would continue to rise towards small $p_{\rm t}$ values and
would give a thermal contribution which is larger than the values of
$r_{\rm B > \Delta}$ quoted in table 2, in agreement with \cite{wein96}.
\begin{figure}
\epsfxsize=8.5cm
\epsffile[15 120 550 420]{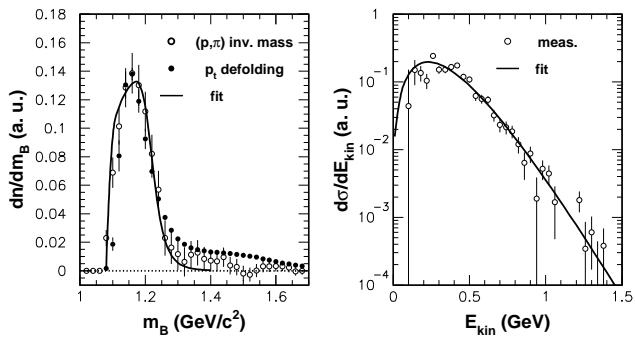}
\caption{Left panel: The measured (symbols) invariant mass spectrum of the Ni 
+ Ni reaction at 1.93 AGeV bombarding energy, and the fit (curve) using the
Boltzmann corrected $f_{\Delta}(m_{\rm B})$ distribution of 
\protect\cite{wein96}. Right panel:
The kinetic energy spectrum 
(symbols) of the $\Delta(1232)$ resonances from the same reaction. 
The curve is a fit employing the Siemens-Rasmussen formula 
\protect\cite{siem79}.
The values for the fitted temperature (left panel) and radial velocity (right 
panel) are listed in table 3}
\end{figure}

With these informations and
the known \cite{fopi1,fopi2} dependence of the pion multiplicity on
the number of participants $A_{\rm part}$, one may calculate the fraction
$n_{\Delta}$ of participants which are excited to $\Delta(1232)$ resonances
under the present experimental conditions. The values of $n_{\Delta}$ are
listed in table 2, they increase with bombarding energy primarily because
the number of pions increases. Similar values for the fraction of
$\Delta(1232)$ resonances populated in the Ni + Ni reactions were extracted
in \cite{fopi5}, where the high-momentum tail was interpreted as due to thermal
pions emitted from an equilibrated radially expanding source with the 
temperatures $T$ and the average expansion velocities $\beta$ listed in table 1.

According to the observed mass shifts in the current analysis
the freeze out conditions for pions
from late $\Delta(1232)$ decays would be
characterized by a much lower value of
$T_{\Delta} = 40 - 60$ MeV. This value when
interpreted as temperature is of similar size as the average freeze out
temperatures of protons in the complete phase space,
extracted in recent analyses of the same present data samples
\cite{fopi6,fopi7} which also point to a larger collective
energy and thus smaller temperatures than estimated from the kinetic energy
spectra of pions and light charged particles alone.
It is also in line with the analysis of
the Au + Au reaction at lower bombarding energies \cite{fopi3}.

On the other
hand, the assumption of thermal equilibrium and instant freeze out at
$T_{\Delta}$ fails to reproduce the $\Delta(1232)$ yield.
As was shown in \cite{fopi5}, where the attainment of equilibrium was
postulated, the considerably higher temperatures listed in table 1
are necessary to describe the observed production probabilities.
These findings are in agreement with \cite{Averbeck98} and are confirmed by
an improved calculation where the widths of all higher-lying
resonances were included via a Breit-Wigner parametrisation with
parameters listed in the particle data booklet \cite{pdb90}.
For a consistent treatment of the yields and
the energy distributions of baryons and pions we drop the surface corrections
and use the phase space density of infinite nuclear matter \cite{dani95}.
At a temperature of $T = 92$ MeV this model
then predicts a value of approximately $r_{\rm B > \Delta} = 0.35$ for the
ratio between higher-lying and the $\Delta(1232)$ resonance, in variance with
the measured value given in table 2.
From the thermal model one
may also deduce the contribution of thermal pions which increases with
decreasing freeze out density. Estimating the upper limit of this contribution
from the $\kappa_{\rm p}^{\pm}$ value of table 1 and the published pion yield
\cite{fopi1,fopi2} the freeze out density $\rho$ should be close to
normal nuclear matter density $\rho_0$, whereas it is predicted that the
majority of pions, i.e. those with small momenta, freeze out at $\rho \approx
\rho_0 / 3$ \cite{teis97b,mahe97}.
\begin{figure*}
\epsfxsize=13.5cm
\epsffile[0 160 560 380]{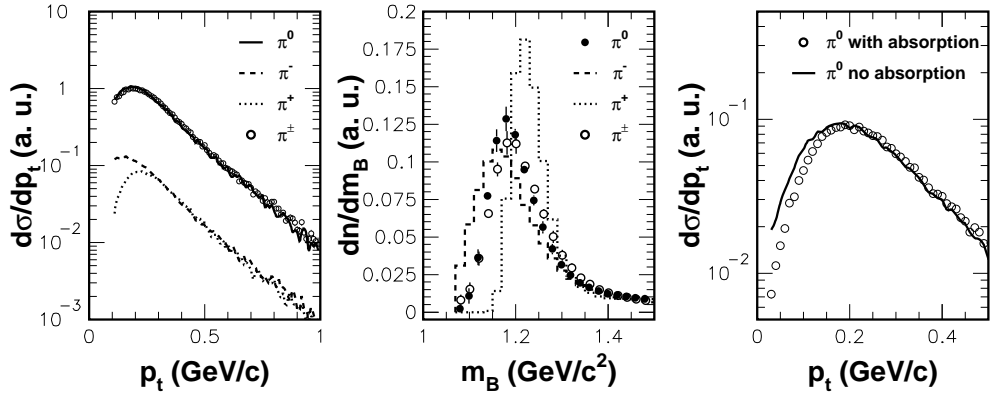}
\caption{The left panel displays the calculated $p_{\rm t}$ distributions of
$\pi^0$, $\pi^-$, and $\pi^+$ (curves). The circles present the
sum of the $\pi^-$ and $\pi^+$ distributions. The centre panel displays
the invariant mass distributions obtained by defolding the $p_{\rm t}$
spectra of the left panel. The right panel illustrates the reduction of
low-momentum $\pi^0$ caused by their absorption in the source }
\end{figure*}

These inconsistencies in the interpretation of the different experimental
observations probably point to the fact that
the simple picture of an instantaneous freeze out in thermal and
chemical equilibrium is not justified.
The number of baryon resonances
and pions is determined very early in the reaction when the energy to excite
these resonances is still sufficiently large \cite{dani95}. The dominant
part of the $\Delta(1232)$ resonances is measured at late stages of the
reaction, when this energy has considerably decreased because of earlier $\pi$
emissions, and because of the development of radial flow. In case one is
allowed to interpret $T_{\Delta}$ as that part of the energy corresponding to
unordered motion at freeze out one may deduce from resonances in the mass range
$m_{\rm B} < m_<$ also the amount of energy converted into flow
at the late stage of the reaction. We have reconstructed for those
resonances their kinetic energy spectrum which is depicted in the right panel
of Fig.8 for the Ni + Ni reaction at 1.93 AGeV bombarding energy.
The reconstruction procedure is equivalent to the one used to identify the
resonances, i.e. the $E - m_{\rm B}$ spectrum of mixed pairs
is subtracted from the one of measured pairs with a ratio $r^{\pm}$ which
was determined earlier and is listed in table 1. The kinetic energy spectra
are corrected for the effects of the $CDC$ geometry onto $E$ and $m_{\rm B}$.
The resulting $E_{\rm kin}$
distribution of the shifted $\Delta(1232)$ resonance appears to have a
maximum close to $E_{\rm kin} = 0.2$ GeV. The shape of this distribution
suggests the existence of a strong collective flow. The fit by means of the
Siemens - Rasmussen formula \cite{siem79} with fixed mass $m_{\Delta} = 1180$
MeV/c$^2$ and temperature $T_{\Delta} = 59$ MeV  and shown as full curve in the
right panel of Fig.8, yields a flow velocity $\beta_{\Delta} = 0.53 \pm 0.04$.
The values of $\beta_{\Delta}$ deduced similarly from the other kinetic energy
spectra of low-lying resonances are listed in table 3. If interpreted in this
way it shows that at late freeze out times of the pions the temperatures do
not change significantly in the Ni + Ni systems, but the radial flow
velocities increase with energy. The Au + Au system, compared with the Ni +
Ni system, is characterized by a lower temperature and larger radial flow
velocity.

Because of the very low nuclear temperatures deduced from our analysis
one cannot neglect the possibility that the shifts of
the resonance mass might be induced by still other effects of the nuclear
environment than those already considered in \cite{wein96}. In \cite{ko96}
the alternatives are discussed in some detail. In case the mass shift
scales with the nuclear density and higher lying resonances are excited
more abundantly in the earlier times of large densities the observed mass
spectrum of the baryon resonances would be rather complex and difficult to
decompose in contributions from specific resonances. From
Fig.6 it appears that whatever the true mechanism is, it produces a resonance
mass distribution with near-exponential decline towards larger mass values.
It was indeed this exponential decline which was used very early by Chapline
et al. \cite{chap73} to study the properties of dense hadronic matter.

\section{Appendix}
\subsection{Effects due to the Coulomb interaction}
The Coulomb residual interaction between the nuclear matter distribution and
charged mesons modifies the primordial phase space distributions of the latter
after they were emitted by resonance decays.
The size of this modification is particularly
strong for pions because of their small mass, and it is readily observed in
experiments: The momentum distribution of positive and negative pions differ
for small momenta, the difference is still observed in systems with total
charge $Z < 60$ \cite{fopi2}. In a recent theoretical paper based on $BUU$
calculations \cite{teis97b}, it was proposed that the $\pi^- / \pi^+$
difference provides a sensitive method to investigate the nuclear matter
distribution at $SIS$ energies.

In our Monte Carlo studies of how the Coulomb effects may change the deduced
mass distributions of baryon resonances we have employed a slightly simpler
model which is based on a spherical and radially expanding baryon distribution
in thermal equilibrium. The $\Delta(1232)$ resonances with the mass distribution
$f_{\Delta}(m_{\rm B})$ are present in the baryon distribution with fraction
$n_{\Delta}$, they can decay via $\pi^+$, $\pi^-$, or $\pi^0$ emission only at 
the surface of this distribution. After the decay the emitted particles are 
propagated in the Coulomb field of the baryon distribution. The positions and 
the momenta are calculated in relativistic kinematics, the time steps are
sufficiently small to handle the complete momentum range of pions. The
final momentum is reached, when the relative change 
of the momentum becomes smaller than $10^{-6}$. In case the total energy
approaches the pion rest mass, this is only possible for $\pi^-$ because of
their negative Coulomb potential, the particle is considered to be in a bound
state and cannot be detected. Furthermore we use an absorption 
probability of 0.2 for those pions, which revert into the 
baryon distribution. Notice that absorption can occur for all pion charges,
the probability depends on the relation between the momentum of the decaying
resonance and the momentum of the pion emitted by the decay.
To be close to the situation of the Au + Au reaction we consider a source
of normal nuclear matter density with 310 baryons and a total charge given
by the $N/Z$ ratio of Au \cite{mahe97}. The kinetic energy of the 
baryons is assumed to be distributed according to the formula of Siemens and
Rasmussen \cite{siem79} with a temperature $T = 81$ MeV and an expansion
velocity $\beta = 0.32$. In addition to the geometrical and the detection
limits of the $CDC$ we also include the finite momentum resolutions of the
$CDC$, characterized by the parameters $\sigma(p_{\rm t}) / p_{\rm t} 
= 0.05$, $\sigma(\varphi) = 1^{\circ}$, and $\sigma(\vartheta) = 5^{\circ}$
\cite{fopi1}.

\begin{figure}
\epsfxsize=8.5cm
\epsffile[10 150 560 440]{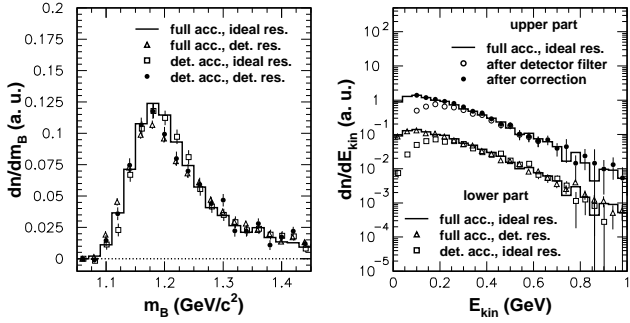}
\caption{ The influences of finite detector acceptance and resolution on the 
resonance mass (left panel) and kinetic energy (right panel)
distributions deduced from correlated $(p, \pi^0)$ pairs}
\end{figure}

This simple model is sufficient to reproduce the gross features of the
measured \cite{fopi1,fopi2} transverse momentum spectra of $\pi^{\pm}$,
i.e. their difference at momenta $p_{\rm t} < 0.3$ GeV/c, and their identical
decline for higher transverse momenta, see Fig.9 left panel. It is also
evident that the sum of the $\pi^+$ and $\pi^-$ momentum distributions is a
good approximation for the primordial momentum distribution obtained from
$\pi^0$. It is therefore not surprising that the primordial $\Delta(1232)$ mass
distribution $f_{\Delta}(m_{\rm B})$ is not retrieved from the distorted pion
momenta. The defolding of the pion $p_{\rm t}$ spectra yields from the
$\pi^-$ respectively $\pi^+$ a $f_{\Delta}(m_{\rm B})$ distribution which is
shifted to smaller/larger masses, see Fig.9 centre panel, whereas the sum of
both $p_{\rm t}$ spectra allows to obtain with good approximation the
undistorted $f_{\Delta}(m_{\rm B})$ distribution as expected since the
Coulomb effects cancel when the $p_{\rm t}$ spectra of $\pi^-$ and
$\pi^+$ are added.

It might be interesting to illustrate how the pion absorption in nuclear
matter changes the pion $p_{\rm t}$ distributions. In Fig.9 right panel the
$p_{\rm t}$ spectra of $\pi^0$ are compared with and without absorption. It is
obvious that absorption mainly suppresses pions with small transverse momenta.
On the other hand this part of the momentum distribution becomes enhanced
again when the decay channel into $2\pi$ opens. Therefore the net effect is
presumably small, and it is beyond the scope of the present study to
disentangle these effects experimentally.

\begin{figure}
\epsfxsize=8.5cm
\epsffile[70 90 440 440]{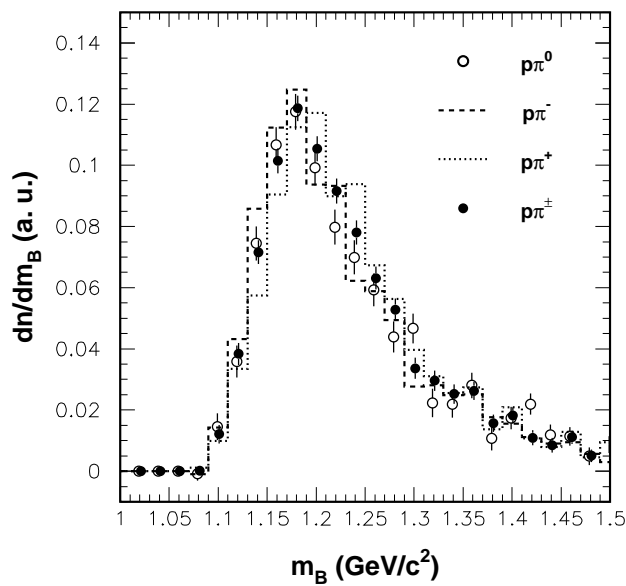}
\caption{Comparison between the resonance mass distributions from
($p\pi^0$), ($p\pi^-$), and ($p\pi^+$) pairs. The ($p\pi^{\pm}$) distribution
corresponds to the normalized sum of the $(p\pi^-)$ and $(p\pi^+)$
distributions}
\end{figure}
The situation becomes more complicated when looking at the correlations
between pions and protons. In Fig.1 we have shown how the geometric boundaries
of the $CDC$ reduce the detection efficiency for small resonance masses. In
the left panel of Fig.10 the uncorrected invariant mass spectrum from
$(p, \pi^0)$ pairs is plotted which shows the size of this effect. On the
other hand, when the finite detector resolutions are considered one finds an
enhancement of the invariant mass distribution at small masses. Including
both effects one obtains in our Monte Carlo simulations an invariant mass
distribution very close to the input. The cancellation of different effects
is not always observed. In case of the energy distribution of resonances we
find in the right panel of Fig.10 that these are mainly effected by the
geometric boundaries, and that corrections have to be applied to restore
the original distribution from the measurement. The corrections are considered
reliable only for kinetic energies $E_{\rm kin} > 100$ MeV.

The effects of the residual Coulomb interaction on the invariant masses from
$(p, \pi^{\pm})$ pairs are shown in Fig.11. Qualitatively the situation is
similar to the analysis of the $p_{\rm t}$ spectra shown in Fig.9.
The $f_{\Delta}(m_{\rm B})$ distribution deduced from the $(p, \pi^-)$
respectively $(p, \pi^+)$ pairs are also shifted to smaller/larger masses
relative to the mass distribution from correlated $(p, \pi^0)$ pairs. On the
other hand the average mass distribution from both $(p, \pi^-)$ and
$(p, \pi^+)$ pairs comes very close to the undistorted distribution.

\begin{table}
\caption{The values of the mass $m_{\Delta}$ and the width $\sigma_{\Delta}$
obtained from $\pi^0$, $\pi^-$, $\pi^+$, and the average $\pi^{\pm}$ by the
analysis of $(p, \pi)$ pairs or by the defolding technique}
\begin{center}
\begin{tabular}{|c|c|c|c|c|}
\hline
 &  \multicolumn{2}{|c|}{pair analysis} & \multicolumn{2}{|c|}{defolding}\\
 & $m_{\Delta}$ & $\sigma_{\Delta}$ & $m_{\Delta}$ & $\sigma_{\Delta}$   \\
 & (MeV/c$^2$)  &    (MeV/c$^2$)    & (MeV/c$^2$)  &    (MeV/c$^2$)      \\
\hline
 $\pi^0$     & 1201 $\pm$ 3 & 51 $\pm$ 2 & 1198 $\pm$ 2 & 48 $\pm$ 1 \\
\hline
 $\pi^-$     & 1197 $\pm$ 3 & 50 $\pm$ 2 & 1180 $\pm$ 3 & 56 $\pm$ 2 \\
\hline
 $\pi^+$     & 1207 $\pm$ 3 & 49 $\pm$ 2 & 1228 $\pm$ 2 & 34 $\pm$ 2 \\
\hline
 $\pi^{\pm}$ & 1202 $\pm$ 2 & 50 $\pm$ 2 & 1202 $\pm$ 2 & 51 $\pm$ 2 \\
\hline
\end{tabular}
\end{center}
\end{table}

The sizes of the mass shifts $m_{\Delta}$ and the widths $\sigma_{\Delta}$
deduced from our Monte Carlo studies are listed in table 4. 
It is evident that the results from the correlation analyses are less
affected by the Coulomb distortion than those from the defolding analysis.
Nevertheless adding the $\pi^-$ and $\pi^+$ data one comes closest to the
primordial conditions, which is the procedure employed in the present analysis.

\subsection{Effects due to the tracking efficiency}
The discussion in section 2 is primarily concerned with the efficiency of
reconstructing the baryon resonances within the geometrical boundaries of the
$CDC$. Another efficiency reduction, which requires a careful investigation,
is imposed by the limited capabilities of the available tracking algorithms
to separate and identify individual particle tracks in the $CDC$. These
inefficiencies become largest when the mass of the decaying resonance is close
to the threshold mass $m_0$, i.e. when the relative momentum between the proton
and the pion is small and the two particle tracks are close to each other. For
the tracking algorithm, which analyzes particle tracks in the $xy$ plane
perpendicular to the $z$ direction of the magnetic field, the closeness of
tracks is given by the relative angle $\alpha_{\rm t} = \varphi^{(\pi)} -
\varphi^{(\rm p)}$ and by the magnitude of $p_{\rm t}^{(\pi)}$. The smaller
$\alpha_{\rm t}$ and the larger $p_{\rm t}^{(\pi)}$ the more difficult it
becomes to separate pion tracks from proton tracks.

The size of the efficiency reduction as function of $\alpha_{\rm t}$ and
$p_{\rm t}^{(\pi)}$ can be obtained from a comparison of the invariant
mass distributions $\frac{dn(meas)}{dm_{\rm B}}$ and
$\frac{dn(mixed)}{dm_{\rm B}}$ as functions of $\alpha_{\rm t}$ and
$p_{\rm t}^{(\pi)}$ since in the latter all pairs are identified with
identical efficiencies. We show in Fig.12 the ratio 
$\omega = \frac{dn(meas)}{dn(mixed)}$
for $(p, \pi^-)$, $(p, \pi^+)$, and Monte Carlo generated $(p, \pi)$ tracks as
functions of $\alpha_{\rm t}$, where the Monte Carlo simulation illustrates the
ideal case of complete tracking efficiency.
In the simulation the conditions are similar to the experimental ones, in 
particular the fraction of $(p, \pi^{\pm})$ pairs due to resonance decay is in
the order of 1\%. The comparison with the Monte Carlo simulation proves,
that the probability to miss
correlated $(p, \pi^{\pm})$ pairs never exceeds 20\% for a given value of
$\alpha_{\rm t}$, and that it is certainly more appropriate to correct for
this inefficiency than to remove all data with $| \alpha_{\rm t} | <
50^{\circ}$, as was done in \cite{trza94}.

\begin{figure}
\epsfxsize=8.5cm
\epsffile[80 90 440 440]{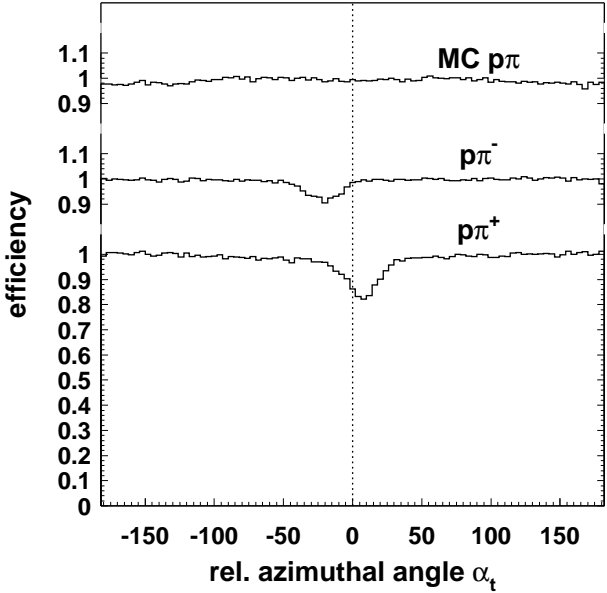}
\caption{ Efficiency to identify a proton and a pion track as function of the
relative angle $\alpha_{\rm t}$ between these two tracks in the $x,y$ plane
perpendicular to the magnetic field}
\end{figure}

That the efficiency reduction is different for $(p, \pi^-)$ and $(p, \pi^+)$
pairs was to be expected since the $p$ and $\pi^+$ have the same curvature
signs, whereas these are different in case of $p$ and $\pi^-$. Furthermore
the fact that in case of $(p, \pi^-)$ pairs the efficiency is reduced only for
negative $\alpha_{\rm t}$ angles offers the possibility to check the consistency
of the efficiency correction. The invariant mass
spectra $\frac{dn}{dm_{\rm B}}$ have to be the same for $\alpha_{\rm t} < 0$
and $\alpha_{\rm t} > 0$. This check was done in our analyses and was shown
to yield identical $\frac{dn}{dm_{\rm B}}$ distributions within the given
uncertainties. All data shown in this paper were therefore efficiency corrected
according to the results of Fig.12 and the similar dependences on
$p_{\rm t}^{(\pi)}$. The sizes of the efficiency corrections were directly
obtained from the data, their dependence on $\alpha_{\rm t}$ and
$p_{\rm t}^{(\pi)}$ was, however, smoothed by the spline algorithm.

\noindent {\bf Acknowledgement}
We would like to thank W. Weinhold for providing us 
with his $f_{\Delta}(m_{\rm B})$ function. This work was supported in part by
the Bundesministerium f\"ur Forschung und Technologie under contract
06 HD 525 I(3) and by the Gesellschaft f\"ur Schwerionenforschung under
contract HD Pel K and by the Korea Research Foundation under contract 
1997-001-D00117.


\begin{thebibliography}{References}
\bibitem{fopi1} {\bf FOPI} collaboration: D. Pelte et al., Z. Phys. A.
 {\bf 357}, 215(1997)
\bibitem{fopi2} {\bf FOPI} collaboration: D. Pelte et al., Z. Phys. A.
 {\bf 359}, 55(1997)
\bibitem{bert88} G.F. Bertsch, S. Das Gupta, Phys. Rep. {\bf 160}, 189(1988)
\bibitem{cass88} W. Cassing, K. Niita, S.J. Wang, Z. Phys.  A {\bf 331},
   439(1988) 
\bibitem{cass90} W. Cassing, V. Metag, U. Mosel, K. Niita, Phys. Rep.
   {\bf 188}, 363(1990)
\bibitem{aich91} J. Aichelin, Phys. Rep. {\bf 202}, 233(1991)
\bibitem{bass93} S.A. Bass GSI-93-13 Report(1993)
\bibitem{bass94} S.A. Bass, C. Hartnack, H. St\"ocker, W. Greiner, Phys. Rev.
   C {\bf 50}, 2167(1994)
\bibitem{bass95} S.A. Bass, C. Hartnack, H. St\"ocker, W. Greiner, Phys. Rev.
   C {\bf 51}, 3343(1995)
\bibitem{dani95} P. Danielewicz, Phys. Rev C {\bf 51}, 716(1995)
\bibitem{teis97a} S. Teis, W. Cassing, M. Effenberger, A. Hombach, U. Mosel,
   G. Wolf, Z. Phys. A. {\bf 356}, 421(1997)
\bibitem{broc84} R. Brockmann, J.W. Harris, A. Sandoval, R. Stock,
   H.Stroebele, G. Odyniec, H.G. Pugh, L.S. Schroeder, R.E. Renfordt,
   D. Schall, D. Bangert, W. Rauch, K.L. Wolf, Phys. Rev. Lett. {\bf 53},
   2012(1984)
\bibitem{sand85} A. Sandoval, R. Brockmann, R. Stock, H.Stroebele,
   D. Bangert, W. Rauch, R.E. Renfordt, D. Schall, J.W. Harris, G. Odyniec,
   H.G. Pugh, K.L. Wolf, GSI-85-10 Report(1985)
\bibitem{seng94} P. Senger, in: Multiparticle Correlations and Nuclear
   Reactions, ed. by J. Aichelin and D. Ardouin (World Scientific Publ. Co.,
   1994), page 285
\bibitem{e814} {\bf E814} collaboration: J. Barrette et al.,
   Phys. Lett. B {\bf 351}, 93(1995)
\bibitem{trza91} {\bf DIOGENE} collaboration: M. Trzaska et al.,
   Z. Phys. A {\bf 340}, 325(1991)
\bibitem{chib91} J. Chiba, T. Kobayashi, T. Nadae, I. Arai, N. Kato,
   H. Katayama, A. Manabe, M. Tanaka, K. Tomizawa, D. Beatty, G. Edwards,
   C. Glashauser, G.J. Kumartzki, R.D. Ransome, T.T. Baker, Phys. Rev. Lett.
   {\bf 67}, 1982(1991)
\bibitem{henn92} T. Hennino, B. Ramstein, D. Bachalier. H.G. Bohlen,
   J.L. Boyard, C. Ellegaard, C. Gaarde, J. Gosset, J.C. Jourdain,
   J.S. Larsen, M.C. Lemaire, D.L. Hote, H.P. Morsch, M. \"Osterlund,
   J. Poitou, P. Radvanyi, M. Roy-Stephan, T. Sams, K. Sneppen, O. Valette,
   P. Zupranski, Phys. Lett. B {\bf 283}, 42(1992)
\bibitem{eos1} {\bf EOS} collaboration: E.L. Hjort et al., Phys. Rev. Lett.
   {\bf 79}, 4345(1997)
\bibitem{trza94} M. Trzaska, in: Multiparticle Correlations and Nuclear
   Reactions, ed. by J. Aichelin and    D. Ardouin (World Scientific Publ.
   Co.,  1994), page 95
\bibitem{berg94} {\bf TAPS} collaboration: F.-D. Berg et al., Phys. Rev.
   Lett. {\bf 72}, 977(1994)
\bibitem{eos2} {\bf EOS} collaboration: M.A. Lisa et al., Phys. Rev. Lett.
   {\bf 75}, 2662(1995)
\bibitem{muen97} {\bf KaoS} collaboration: C. M\"untz et al., Z. Phys.
   A {\bf 357}, 1399(1997)
\bibitem{siem79} P.J. Siemens, J.O. Rasmussen, Phys. Rev. Lett. {\bf 42},
   880(1979)
\bibitem{soll91} J. Sollfrank, P. Koch, U. Heinz, Z. Phys. C {\bf 52},
   593(1991)
\bibitem{demp77} A.P. Dempster, N.M. Laird, D.B. Rubin, J. Roy. Stat. Soc. B
   {\bf 39}, 1(1977)
\bibitem{durs95} H. Durst, Staatsexamensarbeit (Faculty of Mathematic,
   University of Heidelberg (1995)), unpublished
\bibitem{fopi5} {\bf FOPI} collaboration: N. Herrmann et al., Nucl. Phys.
   {\bf A610}, 49c(1996); B. Hong, et al., Phys. Lett. B {\bf 407}, 115(1997);
   B. Hong et al., Phys. Rev. C {\bf 57}, 244(1998)
\bibitem{hote94} D. L'H\^ote, Nucl. Instr. Meth. {\bf A337}, 544(1994)
\bibitem{fopi6} {\bf FOPI} collaboration: M. Korolija et al., to be published
\bibitem{cugn88} J. Cugnon, M.C. Lemaire, Nucl. Phys. {\bf A489}, 781(1988)
\bibitem{ko96} C.M. Ko, G.Q. Li, J. Phys. G {\bf 22}, 1673(1996)
\bibitem{kaos1} {\bf KaoS} collaboration: R. Barth et al., Phys. Rev. Lett.
   {\bf 78}, 4007(1997)
\bibitem{fopi4} {\bf FOPI} collaboration: J. Ritman et al., Z. Phys. A
   {\bf 352}, 355(1995)
\bibitem{wein96} W. Weinhold, B.L. Friman, W. N\"orenberg, Acta Phys. Pol. B
   {\bf 27}, 3249(1996)
\bibitem{fopi7} {\bf FOPI} collaboration: R. Kotte et al., Z. Phys. A
   {\bf 359}, 47(1997)
\bibitem{fopi3} {\bf FOPI} collaboration: W. Reisdorf et al., Nucl. Phys.
   {\bf A612}, 493(1997)
\bibitem{Averbeck98} R. Averbeck et al., nucl-ex/9803001, to be published
\bibitem{pdb90} Rev. of Part. Prop., Phys. Lett. Phys. Lett. B {\bf 239} (1990)
\bibitem{teis97b} S. Teis, W. Cassing, M. Effenberger, A. Hombach, U. Mosel,
   G. Wolf, Z. Phys. A {\bf 359}, 297(1997)
\bibitem{mahe97} V.S. Uma Maheswari, C. Fuchs, A. Faessler, L. Sehn, D.S. Kosov,
   Z. Wang, nucl-th/9706004, to be published in Nucl. Phys. {\bf A}
\bibitem{chap73} G.F. Chapline, M.H. Johnson, E. Teller, M.S.Weiss, Phys. Rev.
   D {\bf 8}, 4302(1973)
\end{thebibliography}
\end{document}